\documentclass[runningheads]{llncs}
\usepackage{amsfonts}
\usepackage{graphicx}

\usepackage{algpseudocode}
\makeatletter
\renewcommand{\ALG@beginalgorithmic}{\tiny}
\makeatother
\algrenewcommand\alglinenumber[1]{\tiny #1:}

\usepackage{amsmath}
\DeclareMathOperator*{\argmax}{arg\,max}

\begin{document}
\title{Learning to Cooperate with Completely Unknown Teammates
\thanks{This work was partially supported by national funds through FCT, Funda\c{c}\~{a}o para a Ci\^{e}ncia e a Tecnologia, under project UIDB/50021/2020 (INESC-ID multi-annual funding), the HOTSPOT project, with reference PTDC/CCI-COM/7203/2020, and the RELEvaNT project, with reference PTDC/CCI-COM/5060/2021. In addition, this work was partially supported by TAILOR, a project funded by EU Horizon 2020 research and innovation programme under GA No 952215.}}
%
%
\author{Alexandre Neves\inst{2} \and Alberto Sardinha\inst{1,2}}
\authorrunning{A. Neves and A. Sardinha}
%
\institute{INESC-ID \and Instituto Superior T\'{e}cnico, Universidade de Lisboa
\email{jose.alberto.sardinha@tecnico.ulisboa.pt}}
\maketitle              
\begin{abstract}
A key goal of ad hoc teamwork is to develop a learning agent that cooperates with unknown teams, without resorting to any pre-coordination protocol. Despite a vast number of ad hoc teamwork algorithms in the literature, most of them cannot address the problem of learning to cooperate with a completely unknown team, unless it learns from scratch. This article presents a novel approach that uses transfer learning alongside the state-of-the-art PLASTIC-Policy to adapt to completely unknown teammates quickly. We test our solution within the Half Field Offense simulator with five different teammates. The teammates were designed independently by developers from different countries and at different times. Our empirical evaluation shows that it is advantageous for an ad hoc agent to leverage its past knowledge when adapting to a new team instead of learning how to cooperate with it from scratch.

\end{abstract}
\section{Introduction}

As robots become more and more ubiquitous in industrial environments, we also start to see them being deployed in other settings, such as homes \cite{IOCCHI2015258} and hospitals \cite{Tasaki2015Robot}. In tasks that require cooperation, robots should coordinate to achieve a common goal. However, achieving efficient cooperation may be a complex endeavor when robots come from different origins. One way to address this problem is to endow a robot with the ability to learn \emph{on the fly} how to cooperate by observing their teammates and environment.

Ad hoc teamwork \cite{stone2010ad} aims to address the problem above and thus design an agent that learns on the fly to adapt to unknown teammates in order to complete teamwork tasks. Furthermore, these agents must be robust to changes, such as adapting to new teammates and different environments.

Several algorithms for ad hoc teamwork have been proposed over the past few years. The state-of-the-art methods assume an agent has a library of teammate models and/or tasks and propose algorithms that choose on the fly the most appropriate models/tasks to cooperate with the unknown team (e.g., \cite{ribeiro21}, \cite{santos21}, and \cite{ribeiro22}). Other notable examples are the ad hoc agent in \cite{melo2016ad}, where each task within the library is represented as a fully cooperative matrix game, and AATEAM \cite{chen2020ad} which     has a library of teammate models that are learned with attention networks. Probably the most famous algorithm for ad hoc teamwork is PLASTIC-Policy \cite{BARRETT2017132}, which also resorts to a library of learned policies and teammate models.  However, all these algorithms fail to adapt efficiently when an unknown team is very different from the models/tasks within the library.

The main contribution of this work is to address this gap in the ad hoc teamwork literature by combining a transfer learning strategy and PLASTIC-Policy, whereby we use the parameter sharing strategy.  Hence, we create a novel ad hoc agent that can efficiently cooperate with an unknown team that differs significantly from the models/tasks within its library.

We also conducted an empirical evaluation in the Half Field Offense simulation domain \cite{Kalyanakrishnan07}, a modified version of the RoboCup Soccer Simulation 2D sub-league. The results show that an ad hoc agent can indeed take advantage of PLASTIC-Policy combined with a transfer learning method. In our experiments, the ad hoc agent quickly adapts to unknown teammates, exhibiting close-to-optimal behavior from the start.

\section{Background}
\label{sec:background}

A Markov decision process (MDP) is a mathematical framework for building sequential decision-making algorithms for agents. Formally, an MDP is a 5-tuple $(S, A, p, R, \gamma)$, where $S$ is a set of states, $A$ is a set of actions, $p$ is the transition probability function for reaching a state $s'$ given that the previous state was $s$ and the action taken was $a$, $R: S \times A \rightarrow \mathbb{R}$ is the reward received by the agent upon taking an action $a$ from state $s$, and $\gamma$ is the discount factor.

Reinforcement Learning (RL) \cite{SUTTON} is a subarea of machine learning that aims to learn an optimal policy of an MDP, when the agent does not know the transition probability function and reward function. The most famous RL method is Q-learning \cite{watkins1992qlearning}. This RL method learns the action-value function directly, $Q_\pi: S \times A \rightarrow \mathbb{R}$, which corresponds to the discounted reward of taking a given action in a given state then following policy $\pi$. Q-learning updates its estimate of the $Q$-value in the following way:

$$
Q(S_t, A_t) \gets (1 - \alpha) Q(S_t, A_t) + \alpha (R_{t+1} + \gamma \max_a Q(S_{t+1}, a))
$$

where $\alpha$ is the learning rate. Hence, the method interpolates, by a factor of $\alpha$, the current estimate for a given state-action pair with the reward received by taking the given action in the given state added to the best expected discounted reward possible from the next state onward. With a discrete action space and state space, Q-learning converges to the optimal $Q$-values (i.e., the $Q$-values of the optimal policy, $Q_{\pi^*}(s,a)$ or $Q^*(s,a)$) if all actions are sampled a large number of times in all states. An optimal policy is then derived by simply querying which action maximizes the optimal $Q$-value for a given state: $\pi^*(s) = \argmax_a Q^*(s, a)$.

Deep Q-Network (DQN) \cite{mnih2015human} is a popular RL method for estimating the optimal $Q$-values with a neural network. An advantage of this approach is that it can handle continuous state spaces in complex environments. DQN approximates the optimal $Q$-values, $Q^*$, by using a neural network of parameters $\theta$: $Q(s,a;\theta) \approx Q^*(s,a)$. At each step, the agent adds a transition $(s,a,r,s')$ to a replay memory buffer, from which batches of transitions are sampled in order to optimize the parameters of the network by minimizing the following loss:

$$
L(\theta_t) = \mathbb{E}_{s,a,r,s' \sim \rho (.)}\left[ (y_t - Q(s,a;\theta_t))^2 \right]  \textrm{where}\,\, y_t = r + \gamma \max_{a'}Q(s',a'; \theta_{t-1})
$$

Here, $y_t$ is called the temporal difference target, and $y_t - Q$ is called the temporal difference error. $\rho$ represents the behavior distribution, the distribution over transitions $(s,a,r,s')$ collected from the environment. The input to a DQN is a state, $s$, where each feature of the state space corresponds to an input node, and the output are the various values of $Q(s,a)$ where each output node corresponds to a different action $a$.

DQNs also resort to \textit{experience replay} to make their updates more stable, by storing each transition $(s,a,r,s')$ in a circular buffer called \textit{the replay buffer} at every time step. Then, instead of using a single transition to compute the loss and back-propagate, a mini-batch of past transitions (randomly sampled) is used instead. This improves the stability of the updates by using uncorrelated transitions in a batch and is called batch learning.

\section{OTLPP - Online Transfer Learning for Plastic Policy}

In this section, we present our algorithm OTLPP that builds on PLASTIC-Policy \cite{BARRETT2017132}. The key aim of this algorithm is to learn to collaborate quickly with unknown teammates that are significantly different from previously encountered teams.

PLASTIC (Planning and Learning to Adapt Swiftly to Teammates to Improve Cooperation) is an algorithm described in \cite{BARRETT2017132}. When an ad hoc agent uses PLASTIC, it observes how its team acts and models that behavior to predict the optimal action to take. PLASTIC-Policy is the policy-based version of PLASTIC, which allows it to work in complex and continuous state space domains.

PLASTIC-Policy maintains a probability distribution over all team policies within its library, representing how likely the current team is to its library's team model/policies. The ad hoc agent builds this library from the previous interaction with teammates. To update the distribution, the agent observes the team while it acts and determines which past team is most likely to be the current one. A limitation of PLASTIC-Policy is that it is only effective on a team that resembles a team within its library, meaning the current team must be similar to one of the previously encountered teams so that it can cooperate adequately. To solve this problem, Section \ref{subsection:OTLPP} presents an algorithm that combines a transfer learning algorithm with PLASTIC-Policy, which is the key contribution of our work. In Section \ref{subsection:plasticmethods}, we present the methods from PLASTIC-Policy that have been adapted to use a DQN.

\subsection{PLASTIC-Policy Methods}
\label{subsection:plasticmethods}

PLASTIC-Policy relies on the \texttt{LearnAboutPriorTeammates} function (Line 1) to build a team model and policy regarding a previously encountered team. In Fig. \ref{alg:LAPT}, the agent gathers data in the form of the tuple $(s,a,r,s')$\footnote{Where $s$ is the original state, $a$ the action taken, $r$ the reward received and $s'$ the resulting state.} as it plays with a given team. With these tuples, it obtains a policy by using a Deep Q-Network, which is an RL algorithm that uses samples from the transitions to approximate the values of the state-action pairs. It also learns a nearest neighbors model from the same gathered data, to be used later by the PLASTIC-Policy algorithm when making predictions about the new team's behavior.

\begin{figure}[ht]
    \fbox{\begin{minipage}{\textwidth}
    \begin{algorithmic}[1]
        \Function{LearnAboutPriorTeammate}{$t$} \Comment $t$: the prior teammate
            \State DQN $\gets$ a newly initialized Deep Q-Network
            \State Data $\gets \emptyset$ 
            \State $s$ $\gets$ the initial state
            \Repeat
                \State $a$ $\gets$ DQN($s$)
                \State Take action $a$ and collect $(s,a,r,s')$
                \State Data $\gets$ Data $\cup \{(s,a,r,s')\}$ 
                \State Add $(s,a,r,s')$ to DQN's replay buffer and perform batch learning on DQN
                \State $s$ $\gets$ $s'$
            \Until{no more transitions are needed}
            \State Derive a policy $\pi$ for $t$ from DQN
            \State Learn a nearest neighbors model $m$ of $t$ using Data
            \State \Return $(\pi, m)$
        \EndFunction\\

        \Function{UpdateBeliefs}{$BehaviorDistr$, $s$, $s'$} \Comment $BehaviorDistr$: probability distribution over possible teammate behaviors, $s$: the previous environment state, $s'$: the new environment state
            \For{$(\pi, m) \in$ BehaviorDistr}
                \State loss $\gets 1 - P(s'|m, s)$
                \State BehaviorDistr($m$) $\gets $BehaviorDistr($m$)$ * (1 - \eta loss)$ 
            \EndFor
            \State Normalize BehaviorDistr
            \State \Return BehaviorDistr
        \EndFunction\\

        \Function{SelectAction}{$BehaviorDistr$, $s$} \Comment $BehaviorDistr$: probability distribution over possible teammate behaviors, $s$: the current environment state
            \State $(\pi, m) = \textrm{argmax}$ BehaviorDistr \Comment select the most likely policy
            \State \Return $\pi(s)$ \Comment return the action for the given state according to that policy
        \EndFunction

    \end{algorithmic}
    \end{minipage}}
    \fbox{\begin{minipage}{\textwidth}
        \abovecaptionskip=0pt
        \caption{Pseudocode for the policy based implementation of the methods \texttt{LearnAboutPriorTeammate}, \texttt{UpdateBeliefs}, and \texttt{SelectAction} (methods taken from \cite{BARRETT2017132} and adapted to use a DQN).}
        \label{alg:LAPT}
    \end{minipage}}
\end{figure}

In \texttt{LearnAboutPriorTeammates}, a Deep Q-Network is  initialized in Line 2 and the algorithm queries the DQN for which action it should take next (Line 6) and executes it, collecting the tuple $(s,a,r,s')$ in Line 7. Line 10 adds the tuple to \texttt{Data} so that the tuples can be used to build the nearest neighbors model in Line 13. In Line 9, the tuple is also added to the DQN's replay buffer, and the DQN then proceeds to perform batch learning, as described in Section \ref{sec:background}. Once no more transitions are deemed necessary, the agent derives a policy from the Deep Q-Network (Line 12), which can be done by simply saving a snapshot of the network's weights and later using \texttt{DQN}($s$) to select an action, just like Line 6.

PLASTIC-Policy keeps a probability distribution over all previously encountered teams within its library. These beliefs are updated at each timestep as the agent gathers more data regarding the current team. The function \texttt{UpdateBeliefs}, in Line 17 of Fig. \ref{alg:LAPT}, presents the pseudocode for updating the beliefs. Line 18 iterates through all team models (i.e., nearest neighbors models) so that, in line 19, it can use the model, along with the original state $s$, to calculate the probability that the resulting state $s'$ is consistent with the current team. This is done by getting $\hat{s} = m(s)$, where $\hat{s}$ is the closest state in $m$ to $s$, for some distance measure\footnote{The only requirement for the distance measure is it to be 0 when $\hat{s} = s$ and close to 0 when $\hat{s}$ and $s$ are considered "similar".}. Then, the corresponding resulting state, $\hat{s'}$ is obtained from $\hat{s}$\footnote{Such association was recorded in \texttt{LearnAboutPriorTeammates} (Fig. \ref{alg:LAPT}, line 10).}. Furthurmore, each state feature is compared between $s'$ and $\hat{s'}$ to calculate the probability, where the difference between the two is due to a noise drawn from a normal distribution. These probabilities are then multiplied together to obtain $P(s'|m,s)$. In line 20, the probability distribution is updated for the current team model by a factor of $1-\eta \textrm{loss}$. The $\eta$ factor is used to attenuate sporadically incorrect predictions that would otherwise bring the probability of the potentially correct team close to 0. Once all beliefs have been updated, the distribution is normalized (Line 22).

While the agent updates its beliefs, it must also act in the environment to gather further data about the current team. Therefore, it must use one of the policies at its disposal to do so. Choosing which policy to use comes down to simply choosing the team that has the highest belief, according to what the agent has seen so far, as shown in function \texttt{SelectAction} (Line 26 of Fig. \ref{alg:LAPT}).

\begin{figure}[t]
    \fbox{\begin{minipage}{\textwidth}
    \tiny
    \begin{algorithmic}[1]
        \Function{TransferKnowledge}{$BehaviorDistr$} \Comment $BehaviorDistr$: probability distribution over all known teams regarding which one is most similar to the current one.
            \State $(\pi, m) \gets$ argmax BehaviorDistr \Comment get the best policy 
            \State DQN$_{source} \gets$ the Deep Q-Network associated with $\pi$
            \State DQN$_{target} \gets$ a new Deep Q-Network
            \State Use Parameter Sharing between DQN$_{source}$ and DQN$_{target}$
            \State \Return DQN$_{target}$
        \EndFunction
    \end{algorithmic}
    \end{minipage}}
    \fbox{\begin{minipage}{\textwidth}
        \abovecaptionskip=0pt
      \caption{Pseudocode for the transfer learning algorithm used in this article.}
        \label{alg:TK}
    \end{minipage}}
\end{figure}

\subsection{Combining Transfer Learning and PLASTIC-Policy}
\label{subsection:OTLPP}

Online Transfer Learning for PLASTIC-Policy (OTLPP) is the key contribution of this work, where the ad hoc agent can collaborate with a new team that is significantly different from the team models and policies within its library. OTLPP can effectively leverage the ad hoc agent's past experiences with different teams to quickly adapt to a new one by combining PLASTIC-Policy with Transfer Learning.

One common technique for transferring knowledge between two neural networks is to share the parameters (or part of them) between the source network and the target network \cite{zhuang2020comprehensive}. This technique is known as Parameter Sharing, and we used it to transfer knowledge between DQNs. There are several strategies in Parameter Sharing, such as transferring the weights of all layers or just the weights from the shallower (closer to the input) layers. Also, the shallower weights may be frozen, which implies learning will only occur in the deeper layers\footnote{This is done for learning stability reasons.}.

The transfer learning algorithm, represented by the function \texttt{TransferKnowledge} in Fig. \ref{alg:TK}, begins by obtaining the policy that corresponds to the team with the highest probability value, in Line 2. Then, in the following line, it derives a Deep Q-Network from that policy\footnote{The policy is actually computed via a Deep Q-Network, so this derivation is straightforward.}, which will be used as the source for the transfer learning algorithm. The target is a newly initialized Deep Q-Network. The final step is to use Parameter Sharing from the source DQN to the target DQN. As mentioned above, all or just a subset of the weights may be transferred, and weight freezing may occur.

\begin{figure}[t]
    \fbox{\begin{minipage}{\textwidth}
    \tiny
    \begin{algorithmic}[1]
        \Function{OTLPP}{$PriorTeammates$, $HandCodedKnowledge$, $BehaviorPrior$} 
        \Comment $PriorTeammates$: past teammates the agent has encountered, $HandCodedKnowledge$: prior knowledge coded by hand, $BehaviorPrior$: prior distribution over the prior knowledge.
            \State
            \State PriorKnowledge $\gets$ HandCodedKnowledge \Comment initialize knowledge from prior teammates
            \For{$t \in$ PriorTeammates}
                \State PriorKnowledge $\gets$ PriorKnowledge $\cup$ \Call{LearnAboutPriorTeammate}{t}
            \EndFor
            \State BehaviorDistr $\gets$ BehaviorPrior \Comment initialize beliefs
            \State
            \State Initialize $s$
            \Repeat \Comment see which team is the most similar to the current one
                \State $a \gets $ \Call{SelectAction}{BehaviorDistr, $s$}
                \State Take action $a$ and observe $r$, $s'$
                \State BehaviorDistr = \Call{UpdateBeliefs}{BehaviorDistr, $s$, $s'$}
                \State $s \gets s'$
            \Until{one team has high enough probability}
            \State
            \State DQN $\gets$ \Call{TransferKnowledge}{BehaviorDistr} \Comment use Transfer Learning to adapt to the new team
            \State
            \State Initialize $s$ once again
            \Repeat \Comment begin learning with the jump start from the transferred knowledge
                \State $a$ $\gets$ DQN($s$)
                \State Take action $a$ and collect $(s,a,r,s')$
                \State Add $(s,a,r,s')$ to DQN's replay buffer and perform batch learning on DQN
                \State $s$ $\gets$ $s'$
            \Until{the agent has learned}
        \EndFunction
    \end{algorithmic}
    \end{minipage}}
    \fbox{\begin{minipage}{\textwidth}
        \abovecaptionskip=0pt
        \caption{Pseudocode for OTLPP algorithm, which combines transfer learning with the PLASTIC-Policy algorithm from \cite{BARRETT2017132}.}
        \label{alg:OTLPP}
    \end{minipage}}
\end{figure}

In Fig. \ref{alg:OTLPP}, we present the OTLPP algorithm, which combines transfer learning with PLASTIC-Policy. In Lines 3-7, and much like the original PLASTIC-Policy algorithm, the agent begins by combining some knowledge that may have been hand-coded with knowledge gathered from previously encountered teams by using the \texttt{LearnAboutPriorTeammate} function (see Fig. \ref{alg:LAPT}). At this point, the ad hoc agent has compiled a library of policies and team models. Therefore, it can reason about the similarities between the current team and each of these past teams.

In Lines 9-15, the ad hoc agent continuously updates its behavior distribution over the past teams with \texttt{UpdateBeliefs} (Fig. \ref{alg:LAPT}), while also selecting the next action to take based on the current values of the distribution with \texttt{SelectAction} (Fig. \ref{alg:LAPT}). Once a certain team reaches some probability threshold of acceptability, indicating that it is indeed the right team, the agent can move on to transferring knowledge (Line 17) from its prior experiences and begin learning a new policy for the new team. In other words, Line 17, obtains a Deep Q-Network whose parameters have been initialized by \texttt{TransferKnowledge} (Fig. \ref{alg:TK}). It can then begin to learn the new policy for the new team, in lines 19-25, with the advantage of not having to start learning from scratch due to the Transfer Learning algorithm.

\section{Experimental Setup}

This section describes our experimental setup to test the OTLPP algorithm within an agent. We used the Half Field Offense (HFO) simulator \cite{Kalyanakrishnan07}, a modified version of the RoboCup Soccer Simulation 2D sub-league. In HFO, an offense team tries to score a goal and the defense team tries to prevent it. In addition, only half of the original field is playable, and a defending agent cannot attack nor vice-versa. In following paragraphs, we present the experimental setup in detail.

\textbf{Test Setting and Agents:}
All tests were conducted in a 2 vs 2 matches in the HFO, where the ad hoc agent plays with one teammate against two opponents. The opponents are two instances of the \texttt{agent2d} and the teammate can be one of the following: \texttt{agent2d}, \texttt{aut}, \texttt{gliders}, \texttt{helios}, from 2013, and \texttt{receptivity}, from the 2019 RoboCup Soccer Simulation 2D sub-league competition\footnote{Agent binaries were downloaded from the following page:\\ https://archive.robocup.info/Soccer/Simulation/2D/binaries/RoboCup/}. What this means is that the opponents will always act according to the \texttt{agent2d} strategy, whereas the teammate can display one of 5 behaviors, according to each of the 5 different types of teammates mentioned above. No behavioral variability was allowed for the opponents, since \textit{ad hoc teamwork} is not concerned with changes in the task (which includes besting the opponents), hence their singular strategy.

\textbf{State Space:}
HFO provides two state spaces on the fly: a low-level one and a high-level one. The low-level state space provides $59 + 9T + 9O$ features, where $T$ is the number of teammates (excluding the agent) and $O$ is the number of opponents. These features include the positions and velocities of all agents on the field and of the ball, the sines and cosines of various angles (e.g., goal opening, agent orientation, velocity vector orientation), and other game state variables. The high-level state space provides a higher-level view of the match, which combines the lower-level features into more meaningful ones, exposing only $12 + 6T + 3O$ features. Of these, the following 12 features were selected as the state space for our agent: X position (the agent's X-position on the field); Y position (the agent's Y-position on the field); orientation; ball X (the ball's X-position on the field); ball Y (the ball's Y-position on the field); goal opening angle; proximity to opponent; teammate's goal opening Angle; proximity from the teammate to opponent; pass opening angle; teammate's X (the X-position of the teammate; teammate's Y (the Y-position of the teammate). We included the agent's position, along with its orientation and the ball's position, for obvious reasons: the agent cannot make the simplest of decisions if it does not know its position and the ball's location. The goal opening angle feature is present so that the agent may decide whether it is worth shooting toward the goal or pass to a teammate, among other scenarios. This is also why we include the proximity to the opponent, the teammate's goal opening and pass angles, and the teammate's position.

\textbf{Action Space:}
Although HFO offers action spaces of different levels of abstraction. We decided to use a new action space that resorted almost entirely to delegating to the high-level actions already provided by HFO and preventing some unintended behavior, as explained below. As a result, the following 11 actions were made available to the agent: shoot; short-dribble; long-dribble; pass-to; no-op (the agent takes no action for 4 time-steps); go-to-ball; go-to-goal; go-to-teammate; go-away-from-teammate; go-to-nearest-opponent; go-away-from-opponent.

\textbf{Reward Function:}
To create a reward function, we use the status of the game after an action has been taken. The following statuses are available to the agent: in-game (the game is still ongoing); goal; captured-by-defense; out-of-bounds; out-of-time; server-Down. Hence, we use PLASTIC-Policy's reward function \cite{BARRETT2017132}, which also used the HFO to evaluate the agent's performance. The reward function is the following:

\begin{equation*}
R(status) = \left\{
        \begin{array}{ll}
            1000 & \quad \textrm{if $status$ is \texttt{Goal}} \\
            -1 & \quad \textrm{if $status$ is \texttt{In-Game}} \\
            -1000 & \quad \textrm{otherwise}\\
            
        \end{array}
    \right.
\end{equation*}

\textbf{Deep Q-Network Parameters:}
The ad hoc agent was first trained using a Deep Q-Network against all 5 different teammate types from scratch. This allowed the agent to have a library of approximately optimal policies for each teammate type. The network has an input layer of $12$ nodes to accommodate each of the $12$ features from the state space, $3$ hidden layers of $512$ nodes, an output layer of $11$ nodes to indicate the estimated q-value of each of the $11$ actions, Rectified linear unit activation functions between all layers, a learning rate of $0.00025$, a replay memory capacity of $2.5 \times 10^5$ transitions (beginning its use when it has at least $12500$ transitions stored), a learning batch size of $64$ transitions, an $\epsilon$-greedy action selection (with $\epsilon$ having a linear decay that begins at $0.8$ and decays to $0.05$ after $100000$ time-steps), a discount factor, $\gamma$, of 0.995, an Adam optimizer \cite{adam}, a transfer rate of $500$ time-steps, and a weight initialization sampled uniformly from $[-\frac{1}{\sqrt{n}}, \frac{1}{\sqrt{n}}]$, where $n$ is the amount of nodes in a given layer.

\textbf{OTLPP Parameters:}
Like PLASTIC-Policy, our algorithm relies on a parameter, $\eta$, to attenuate sporadic losses on the correct team model. Also, the noise distribution used to compute $P(s'|m,s)$ is a normal distribution of mean 0 and variance $\sigma^2$, so $\sigma$ is also another parameter. The following values were used in our code: $\eta = 0.10$, and $\sigma = 4.0$. The condition \texttt{one team has high enough probability} in Line 15 of Fig. \ref{alg:OTLPP} is true at the end of the 25th game, given that at this stage, the agent almost always makes the correct decision regarding its current team. In the \texttt{UpdateBeliefs} function (Fig. \ref{alg:LAPT}), the probability $P(s'|m,s)$ is calculated by first obtaining the state $\hat{s}$ that is closest to $s$ via the nearest-neighbor model $m$, obtaining that state's corresponding next state, $\hat{s}'$, and then calculating the distance between the next state and the predicted next state, $d = \texttt{Distance}(s', \hat{s}')$. \texttt{Distance} can be defined in several different ways, but fundamentally it should be $0$ if $\hat{s}' = s'$ and close to $0$ if $\hat{s}'$ and $s'$ are considered similar. For this work, and since the intent is to identify which teammate the agent is playing with, the features of the state space that do not pertain to the teammate, such as the opponents' movements or the agent's own movements, should not be considered. Therefore, the only features considered are the teammate's coordinates. Hence, let $s_x$ be the teammate's $x$-position in a given state $s$, and $s_y$ its $y$-position analog. Then, \texttt{Distance} is defined as $\texttt{Distance}(s', \hat{s}') = \prod_{c \in \{x, y\}} \texttt{ProbFromNoise}(s'_c - \hat{s}'_c)
$ and $
\texttt{ProbFromNoise}(\Delta) = 1 - 2 \cdot |F_{N(0, \sigma^2)}(\Delta) - \frac{1}{2}| $, where $F_{N(0, \sigma^2)}$ denotes the cumulative distribution function for a normal distribution of mean $\mu = 0$ and a variance of $\sigma^2$. Therefore, \texttt{ProbFromNoise} returns the probability that a given value was drawn from a normal distribution. As mentioned before, we use a normal distribution in the equation above because we assume that every transition is affected by the noise of the normal distribution.

\textbf{Parameter Sharing:}
During the Transfer Learning stage of the OTLPP algorithm (Fig. \ref{alg:OTLPP}), the agent uses Parameter Sharing to transfer knowledge between a single source Deep Q-Network and the target Deep Q-Network that will later learn the new policy. We decided to transfer all weights, including those from deeper layers, because the environment and the task remain the same. Therefore, a lot of commonality is expected between how the ad hoc agent should behave when paired with the old teammate and with the new one. However, if the change in teams represents a higher change in the overall problem, perhaps transferring only the shallower layers, which correspond to broader abstractions, could also be used.

\section{Results}

In our empirical evaluation, the task consists of having our ad hoc agent (with the OTLPP algorithm) cooperate with one of 5 different teammates, namely (\texttt{agent2d}, \texttt{aut}, \texttt{gliders}, \texttt{helios}, and \texttt{receptivity}). Each teammate was designed independently by developers from different countries for the 2013 and 2019 RoboCup Soccer Simulation 2D sub-league competition. Furthermore, the goal is to score as many goals as possible against a defense team made up of two agents of the type \texttt{agent2d}.

Given that our agent uses a DQN, one way to evaluate the learning procedure at a given point in time is to take a snapshot of its network's parameters at that time and use them to obtain the goal fraction (i.e., the number of goals divided by the number of games). In our experiments, all goal fractions were obtained by running 200 games for the same network parameters and dividing the number of goals scored by that amount. Other metrics may be used, such as the sum of the rewards or the average time to score. However, since the task is ultimately to score a goal, the goal fraction is probably the best metric.

\begin{figure}[h]
    \centering
    \includegraphics[width=1.0\textwidth]{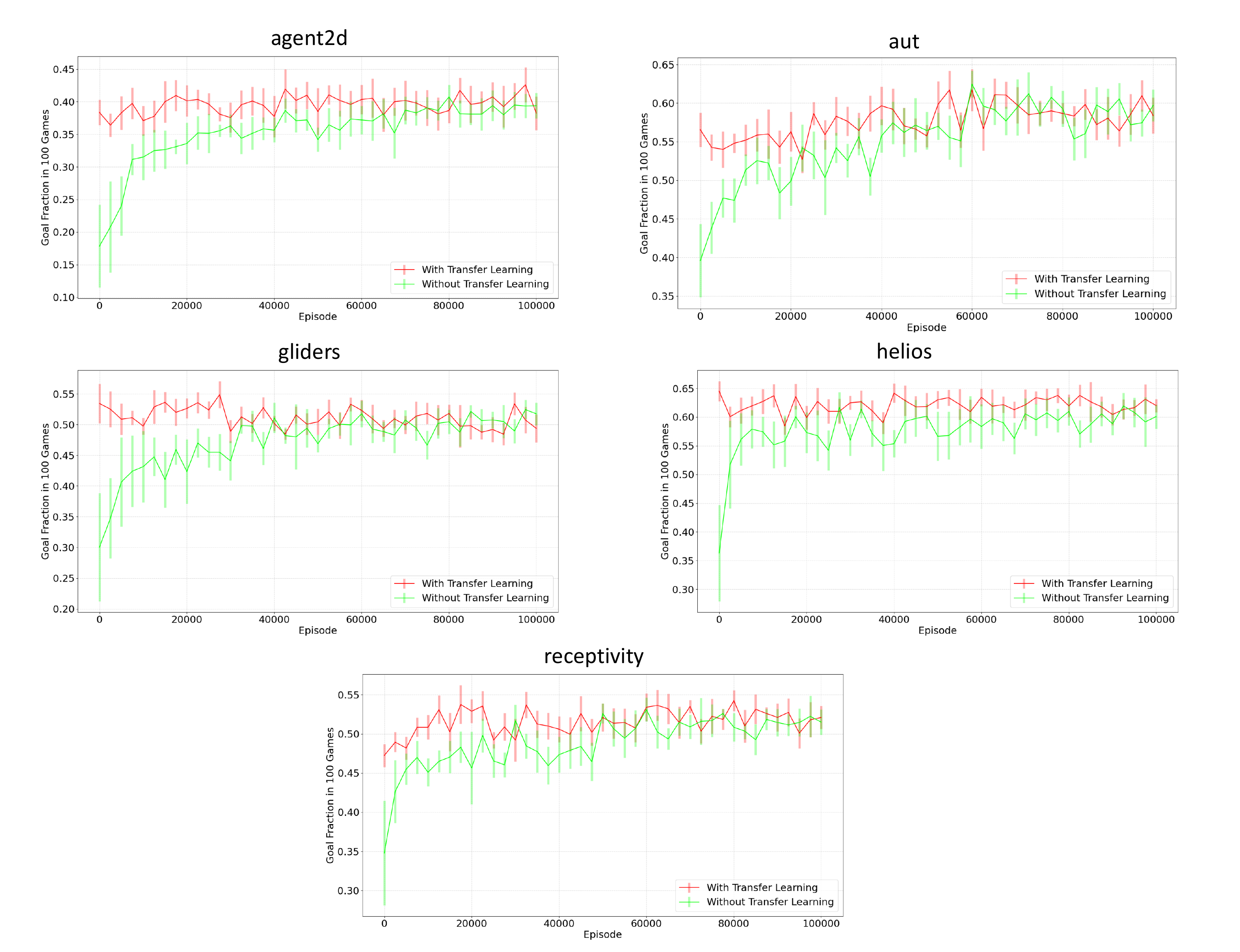}
    \caption{Comparison between the average goal fraction when the agent uses OTLPP (in red) compared to an RL agent (in green), while playing with each teammate. The vertical bars represent one standard deviation.}
    \label{fig:results-learning}
\end{figure}

To test the algorithm against each teammate available, we excluded the policy of the teammate from the library, which is the equivalent to calling \texttt{LearnAboutPriorTeammates} (Fig. \ref{alg:LAPT}) on all teammates except for the particular one during the execution of OTLPP (Fig. \ref{alg:OTLPP}). Hence, the ad hoc agent learns the policy of 4 teammates and then uses that knowledge, along with the behavior distribution it maintains over them, to transfer knowledge to a new Deep Q-Network for the 5\textsuperscript{th} teammate.

In Fig. \ref{fig:results-learning}, we start showing the ad hoc agent's performance during the learning process, when it resorts to the OTLPP algorithm (red line) with each teammate: \texttt{agent2d}, \texttt{aut}, \texttt{gliders}, \texttt{helios}, and \texttt{receptivity}. We also include the performance of an RL agent (green line), which depicts the performance of  an agent that learns from scratch. From the plots, we can notice that our ad hoc agent with the the OTLPP algorithm benefits from a significant boost, when transfer learning takes place at the beginning of learning. In other words, our agent is able to exhibit close-to-optimal behavior from the start.

\begin{table}[h]
\centering
\small
\begin{tabular}{|l|l|l|}
\hline
Team & PLASTIC-Policy & OTLPP \\
\hline
\hline
\texttt{agent2d} & 0.37 & \textbf{0.40} \\
\texttt{helios} & 0.61 & \textbf{0.62} \\
\texttt{aut} & 0.55 & \textbf{0.58} \\
\texttt{gliders} & \textbf{0.53} & 0.51 \\
\texttt{receptivity} & 0.47 & \textbf{0.52} \\
\hline
\end{tabular}
\caption{Comparison between the average goal fraction over 100000 episodes when the agent uses PLASTIC-Policy and when it uses OTLPP, for each team.}
\label{fig:table}
\end{table}

Finally, Table \ref{fig:table} compares OTLPP's performance with PLASTIC-Policy during run time. In particular, Table \ref{fig:table} compares the average score of an ad hoc agent using PLASTIC-Policy with an ad hoc agent using OTLPP, over 100000 episodes. Each ad hoc agent has 4 policies within its library, and we exclude the policy that corresponds to the current teammate of the ad hoc agent. This evaluation enables us to test the performance of the ad hoc agent with a completely unknown teammate. In bold, we highlight the maximum score between both alternatives. The results show that OTLPP can outperform or perform close to PLASTIC-Policy in every test, showcasing the advantage of our OTLPP algorithm. Hence, OTLPP almost always means an improvement over PLASTIC-Policy.

\section{Concluding Remarks}

Our paper is the first work to combine a transfer learning method with PLASTIC-Policy in order to create an ad hoc agent that learns to cooperate with a completely unknown teammate. In our empirical evaluation, we show that the combination of a transfer learning method with PLASTIC-Policy can serve as a powerful tool for ad hoc agents. In particular, the results show that OTLPP can outperform or perform close to PLASTIC-Policy. Hence, the ad hoc agent quickly adapts to completely unknown teammates, exhibiting close-to-optimal behavior from the start. As future work, other transfer learning methods may be explored, such as transferring knowledge from multiple sources \cite{10.1145/2505515.2505603}, meaning multiple teams would be a source of transferred knowledge.

%
%
%
\bibliographystyle{splncs04}
\bibliography{references}

\end{document}